\scrollmode
\documentclass[12pt]{article}
\usepackage{amssymb}
\def\R{{\Bbb R}}

\def\N{{\Bbb N}}
\def\eas{\begin{eqnarray*}}
\def\eeas{\end{eqnarray*}}

\def\nn{\nonumber}
\def\eq#1{\begin{equation}\label{#1}}
\def\eeq{\end{equation}}
\def\ea#1{\begin{eqnarray}\label{#1}}
\def\eea{\end{eqnarray}}
\def\la{\label}
\def\re#1{(\ref{#1})}

\def\ad{\mathop{\mbox{\rm ad}}}

\def\Ker{\mathop{\mbox{\rm Ker}}}

\def\a{{\alpha}}
\def\b{{\beta}}
\def\g{{\gamma}}
\def\d{\partial}

\def\l{{\lambda}}

\def\HH{{\cal H}}

\def\qed{\hfill$\Box$\\\strut}

\def\<{\langle}
\def\>{\rangle}
\newtheorem{lemma}{Lemma}[section]

\newtheorem{corollary}[lemma]{Corollary}
\newtheorem{theorem}[lemma]{Theorem}

\newtheorem{proposition}[lemma]{Proposition}
\def\bib#1{\bibitem[#1]{#1}}
\def\gg{{\mathfrak g}}
\def\kk{{\mathfrak k}}

\def\tf#1#2{{\textstyle\frac{#1}{#2}}}

\def\Aut{\mathop{\mbox{\rm Aut}}}

\def\aut{\mathop{\mathfrak{aut}}}
\def\GG{{\mathcal G}}
\def\gg{{\mathfrak g}}
\def\hh{{\mathfrak h}}
\def\ll{{\mathfrak l}}  
\def\au{\underline{\a}}
\def\bu{\underline{\b}}
\def\tu{\underline{t}}
\def\su{\underline{\tau}}

\def\ecke{\mathbin{\lrcorner}}

\def\XXint#1#2#3{{\setbox0=\hbox{$#1{#2#3}{\int}$}
                 \vcenter{\hbox{$#2#3$}}\kern-.5\wd0}}

\begin{document}
\sloppy
\title{Canonical gauges in higher gauge theory}
\author{Andreas Gastel}
\maketitle

{\abstract \noindent
  We study the problem of finding good gauges for connections in higher
  gauge theories. We find that, for $2$-connections in strict $2$-gauge
  theory and $3$-connections in $3$-gauge theory, there are local
  ``Coulomb gauges''
  that are much more canonical than in classical gauge theory. In particular,
  they are essentially unique, and
  no smallness of curvature is needed in the critical dimensions.
  We give natural definitions of $2$-Yang-Mills and $3$-Yang-Mills theory
  and find that the choice of good gauges makes them essentially linear.
  As an application, (anti-)selfdual $2$-connections over
  $B^6$ are always
  $2$-Yang-Mills, and (anti-)selfdual $3$-connections over $B^8$ are
  always $3$-Yang-Mills.\\[6mm]
{\bf AMS classification. }58E15, 53C08.}

\section{Introduction}

An aspect of gauge theory that has proven to be very important in geometry and
topology is the control of Sobolev norms by the Yang-Mills functional.

Assume that we are working on a trivial principal fibre bundle $B^m\times G$,
where $B^m$ is the unit ball in $\R^m$ and $G$ is a compact Lie group.
Assume $\gg$ is the Lie algebra of $G$, equipped with an $\ad_G$-invariant
scalar product. A connection of the bundle can be described as the
differential operator $d_A$ on $\gg$-valued functions acting as
$d_AX:=dX+[A,X]$. Here $A$ is a $\gg$-valued $1$-form.
This $d_A$ induces similar operators on $\gg$-valued
$k$-forms also denoted by $d_A$. One of the important aspects
of such connections is how they transform under pointwise coordinate
transformations of
the fibres. (This may look unnecessary for the trivial bundle, but it becomes
essential when we start to glue trivial bundles to get nontrivial ones.)
Assume we have a field of transformations $\ad_{g(x)^{-1}}$ acting on
the fibres $\{x\}\times\gg$ of $B^m\times\gg$, which defines a mapping
$g:B^m\to G$. Assume that $g$ is $C^1$, say, and that we want to know what
our $d_A$ looks like after we have applied the coordinate change $\ad_{g(x)^{-1}}$
on all of our fibres. It turns out that the connection $d_A$ is transformed
to $d_{g^*A}$, where $g^*A:=g^{-1}dg+g^{-1}Ag$. This is called a
{\em gauge transformation\/} of $A$.

Observe that the (sufficiently regular) maps $g:B^m\to G$ form the
{\em group of gauge transformations\/} acting from the right on the space
of connections. That the action is from the right is not essential and
a matter of convention. The important point of gauge theory is that
there are quantities derived from $A$ that transform more naturally under gauge
transformations than $A$ itself, i.e.\ like a tensor instead of a differential
operator. The most important such quantity is the {\em curvature\/}
\[
  F_A:=dA+A\wedge A=dA+\frac12\,[A,A]
\]
of $A$, a $\gg$-valued $2$-form that simply transforms as
$F_{g^*A}=g^{-1}F_Ag$. This implies
$|F_{g^*A}|=|F_A|$ since the norm on $\gg$ is $\ad_G$-invariant.
Hence the {\em Yang-Mills functional\/}
\[
  Y\!M(A):=\frac12\int_{B^m}|F_A|^2\,dx
\]
does not change if we transform $A$ by any gauge transformation. It is
therefore a very natural functional to consider. As this is well-known, we
do not go into details.

Given any connection $A$, natural norms like the $W^{1,2}$-norm of $A$
are not gauge-invariant and hence depend on more than only
``the geometric properties of $A$''. Maybe $\|A\|_{W^{1,2}(B^m)}$ is rather
large, but only because we look at $A$ in an unfortunately chosen gauge.
Can we find a gauge transformation such that $\|g^*A\|_{W^{1,2}(B^m)}$ is
controlled by $\|F_{g^*A}\|_{L^2(B^m)}=\|F_A\|_{L^2(B^m)}$? The answer is yes
if $m\le4$ and the $L^2$-norm of the curvature is small enough. This is
Uhlenbeck's \cite{Uh} famous gauge theorem, which is one of the most important
result in gauge theory. In dimensions $m\ge5$, one can still control
$\|g^*A\|_{W^{1,m/2}(B^m)}$ by $\|F_{g^*A}\|_{L^{m/2}(B^m)}$ after a suitable
gauge transformation if $\|F_A\|_{L^{m/2}(B^m)}$ is sufficiently small.
There are more global versions of this, but for us the following local
version will be sufficient.

\begin{theorem}[Uhlenbeck's gauge theorem, \cite{Uh}]\la{thm:Uhl}
  Assume we are given a compact Lie group $G$. Assume
  $A\in W^{1,p}\Lambda^1(B^m,\gg)$ for some $p\ge4$ if $2\le m\le4$, or
  $p\ge\frac{m}2$ if $m\ge5$, represents some connection on the
  trivial bundle $B^m\times G$. There are a constants $\kappa>0$ and
  $c<\infty$ depending
  on $m$ and $p$ only such that, whenever $\|F_A\|_{L^p(B^m)}\le\kappa$, 
  there is a gauge transformation $g\in W^{2,p}(B^m,G)$ such that the
  transformed connection $A':=g^*A$ fulfills $d^*A'=0$, $A'_N=0$ on
  $\d B^m$, and
  \[  \|A'\|_{W^{1,p}(B^m)}\le c\|F_{A'}\|_{L^p(B^m)}.  \]
  The gauge transformation can be estimated by
  \[  \|dg\|_{W^{1,p}(B^m)}\le c\|A\|_{W^{1,p}(B^m)}.   \]
\end{theorem}

Hence the Yang-Mills functional on $4$-dimensional manifolds (where it is
also conformally invariant) locally controls Sobolev norms. Our paper
will be concerned with the question whether there is something similar
for {\em higher gauge theories.}

Higher gauge theories have evolved from attempts to deal with questions
that involve parallel transport not only of vectors (``point locations''),
which is what connections are made for, but also of higher-dimensional
objects. In string theory, the notion of parallel transport of strings
should be useful, and in $M$-theory, it could help to do likewise
with $2$-dimensional branes. In a rich interplay between ideas from
physics and from higher category theory, several higher gauge theories
have evolved, among them the (strict) $2$-gauge theory and $3$-gauge
theory that
we study in this paper. We cannot even attempt to summarize the rich
history that has led to these ideas, and we refer to Baez' and Huerta's
paper \cite{BH} for an excellent overview and an introduction of $2$-gauge
theory. An important step towards $2$-gauge theory was a study of nonabelian
gerbes by Breen and Messing \cite{BM}. For $2$-gauge theory, the reader may
also wish to consult foundational papers by Bartels \cite{Bar}
and by Baez and Schreiber
\cite{BS} (as well as much more work by Baez and/or Schreiber).
For $3$-gauge theory, we refer to a paper by S\"amann and Wolf
\cite{SW} where the theory has been developed, and by Wang \cite{Wa}.

Very roughly, $2$-gauge theory is about $2$-connections on $2$-bundles.
A trivial principal $2$-bundle is described by several data that form a
structure known as a {\em Lie crossed module. }We need two
Lie groups $G$ and $H$ and homomorphisms $t:H\to G$ and $\a:G\to\Aut(H)$
satisfying certain relations. A $2$-connection is described by a
$\gg$-valued $1$-form $A$ and an $\hh$-valued $2$-form $B$, related
to each other by $\tu(B)=F_A$, where here $\tu$ is the differential
of $t$ at $e\in H$. There is a natural $\hh$-valued $3$-form
\[
  Z_{A,B}:=dB+\au(A)\wedge B
\]
which again transforms naturally under $2$-gauge transformations. The latter
are given by a pair $(g,\chi)$ of a function $g:B^m\to G$ and an $\hh$-valued
$1$-form $\chi$. They also form a group acting from the right on the
space of $2$-connections. We will give precise formulae in Section
\ref{Sec:3}. The $L^2$-norm of $Z_{A,B}$ turns out to be
invariant under all $2$-gauge transformations, just as $\|F_A\|_{L^2}$
was invariant under gauge transformations (but not under all
$2$-gauge transformations). Therefore we may reasonably hope that
the $L^2$-norm of $Z_{A,B}$ plays a role in $2$-gauge theory that is
similar to the role of the Yang-Mills functional in gauge theory. We expect
it to be particularly natural in $6$ dimensions, where it is also
conformally invariant.
We therefore call
\[
  Y\!M_2(A,B):=\int_{B^m}|Z_{A,B}|^2\,dx
\]
the {\em $2$-Yang-Mills functional. }The attempt to provide a good notion
of $2$-Yang-Mills has already been undertaken in 2002 by Baez in the
preprint \cite{Bae}. Back then, the significance of the condition
$\tu(B)=F_A$ had not yet been fully established in the theory, hence Baez
works without that condition and considers the functional
$\int(|Z_{A,B}|^2+|F_A-\tu(B)|^2)\,dx$ instead. This is quite natural
in the theory without $\tu(B)=F_A$. In particular, it is also gauge
invariant, but no longer conformally invariant in any dimension. Nevertheless,
there are some interesting aspects in that paper, including some notion of
self-duality in five (!) dimensions.

Our focus is different here, since we have a $2$-Yang-Mills functional that
really resembles Yang-Mills. Therefore, it is tempting to ask whether there
is a higher form of Uhlenbeck's gauge theorem, controlling norms
like $\|A\|_{W^{2,2}}+\|B\|_{W^{1,2}}$ by $\|Z_{A,B}\|_{L^2}$ in dimensions $m\le6$
once a suitable $2$-gauge is fixed,
maybe under a smallness condition for the latter norm. One of our results will
be that this really works. But, surprisingly enough for the author, it
turns out that the good $2$-gauge exists without a smallness condition,
and moreover the transformed $2$-connection has a canonical form that
has the potential to simplify the theory very much. More precisely, we
can $2$-gauge transform $(A,B)$ to get some $(A',B')$ where $A'=0$,
$d^*B'=0$, and $B'$ takes its values in the abelian subalgebra $\Ker\tu$
of $\hh$, plus of course the estimates mentioned above. We call this the
{\em canonical $2$-gauge\/} for $(A,B)$, and we prove that it is even
unique up to a constant gauge transformation. See Section \ref{Sec:4}
for details.

The proof of the existence
of the canonical $2$-gauge is considerably simpler than that of Uhlenbeck's
theorem. Not surprisingly so, since we use the latter --- but
only for connections with $F_A=0$, for which Uhlenbeck's theorem
might look a bit trivial (but actually, under the weak regularity assumptions,
it isn't). The other important ingredient in the proof is Hodge decomposition
on manifolds with boundary.

The existence of a canonical gauge for which $A'$ vanishes and $B'$ maps
to the abelian subalgebra
$\Ker\tu$ makes $2$-Yang-Mills theory an essentially {\em linear\/}
theory, since $Z_{0,B'}=dB'$, and the $2$-Yang-Mills equation becomes the
Laplace equation for $B'$. Is this good news or bad news? On one hand,
$2$-gauge theory is a natural theory that has a geometric content in
describing parallel transport of $1$-dimensional objects, hence we should be
happy to find that the theory turns out to be easier than classical
gauge theory. On the other hand, a theory that is not genuinely nonlinear
may be not the best candidate for a theory with interesting topological
implications like Yang-Mills. And the author suspects that a bit less
linearity would also be expected in physics.

Of course, the ``essentially linear nature'' of $2$-gauge theory has been
remarked before, e.g.\ in the introduction of \cite{SW},
observing that $\tu(Z_{A,B})=0$ follows from Bianchi type
identities, which means the $2$-curvature is always in the ``abelian''
part of the theory.
Also, it has been remarked that there are no examples
of solitons that are ``non-abelian''. Our result makes precise in which sense
the theory is ``linear''. It says that after a canonical
$2$-gauge transformation, we can always work in abelian Lie subalgebras,
where the Euler-Lagrange equations for curvature $L^2$-integrals are linear.
The $2$-gauge transformation itself, however, depends of course
on the $2$-connection and solves a nonlinear system of differential
equations.

One way around having a $2$-gauge theory that is ``too abelian'' may be
to ``embed'' it into $3$-gauge theory where the relations of $2$-gauge theory
do not hold strictly. (This can be given a precise sense in the framework
of categorification, see \cite{SW}.) In $3$-gauge theory, there is a third
Lie group $L$ involved, $3$-connections additionally depend on an
$\ll$-valued $3$-form $C$, and $3$-gauge transformations on an additional
$\ll$-valued $2$-form $\lambda$. There is a curvature $4$-form $Y_{A,B,C}$,
the $L^2$ of which is gauge-invariant, and conformally invariant in $8$
dimensions. We can ask the same question as for $2$-gauge theory.
For $m\le8$, we find canonical gauges where $A'=0$, $B'=0$,
and this time $C'$ is in some abelian Lie subalgebra of $\ll$.
The setting of $3$-gauge theory will be described in Section \ref{Sec:5},
and our gauge theorem in Section \ref{Sec:6}.

One of the points \cite{SW} made in introducing $3$-gauge theory is that
in its framework the curvature $3$-form $Z_{A,B}$ no longer is restricted to
some abelian Lie algebra. The curvature $4$-form $Y_{A,B,C}$, however, does
have that restriction, and this is what makes our results on $3$-gauges
very similar to the ones for $2$-gauges. Note also that $Z_{A,B}$ is not
$3$-gauge covariant in $3$-gauge theory, just as $F_A$ is not $2$-gauge
covariant in $2$-gauge theory.

Our gauge theorems can be applied to {\em flat\/} $2$-connections satisfying
$Z_{A,B}\equiv0$, which then turn out to be $2$-gauge equivalent to the trivial
connection $(0,0)$, and similarly to flat $3$-connections ($Y_{A,B,C}\equiv0$),
which are seen to be $3$-gauge equivalent to $(0,0,0)$. These two statements
can be seen as higher generalizations of the Poincar\'e lemma and have
been proven by Demessie and S\"amann \cite[Thm.\ 2.7 and Thm.\ 2.12]{DS}.
The results are special cases of our gauge theorems, and they are proven
here under rather weak regularity assumptions.

The curious reader may wonder about $4$-gauge theory, but it
has not been defined yet. The problem is that the algebraic
effort to describe $k$-gauge theory seems to grow quickly with $k$.
More precisely, it is the number of mappings and relations needed to describe
a $(k-1)$-crossed module that grows quickly. The notion of a $3$-crossed module,
which should be the basis for $4$-gauge theory, has been developed in
\cite{AKU}. Anyway, the author is not aware of any reason in physics to
consider a $4$-gauge theory.

There are more aspects of gauge theory that turn out to be shared by
$2$-gauge and $3$-gauge theory. In $4$ dimensions, a connection $A$ is called
{\em selfdual or anti-selfdual\/} if ${*}F_A=\pm F_A$. One of the basic
facts in gauge theory is that every (anti-)selfdual connection is
Yang-Mills, i.e.\ it solves the Euler-Lagrange equation $d_A^*F_A=0$
for $Y\!M$. Using our canonical gauges, we prove that in $6$ dimensions
every $2$-connection $(A,B)$ with ${*}Z_{A,B}=\pm Z_{A,B}$ solves the
Euler-Lagrange equation for $Y\!M_2$. A similar result holds for
$3$-connections in $8$ dimensions. We provide details in
the Corollaries \ref{2-Kor} and \ref{3-Kor} below.

\section{Preliminaries on differential forms}

We will need two nontrivial ingredients about differential forms in our study,
the Hopf decomposition with boundaries and Gaffney's inequality.
In preparation of these, we recall that there are two useful forms
of boundary conditions for differential forms. If $M$ is a Riemannian manifold
with smooth boundary, then we can choose coordinates in the neighborhood $V$
of any boundary point $y$ such that $dx_n$ is the dual of the outer normal
on $\d M\cap V$. Let $m:=\dim M$ and $0\le k\le m$. Any $k$-form $\omega$
on $V$ can be decomposed as $\omega=\omega_T+\omega_N$, where
\eas
\omega_T&:=&\sum_{1\le i_1<\ldots<i_k\le n-1}\omega_{i_1,\ldots,i_k}\,dx_{i_1}\wedge
\ldots \wedge dx_{i_k},\\
\omega_N&:=&\sum_{1\le i_1<\ldots<i_k=n}\omega_{i_1,\ldots,i_k}\,dx_{i_1}\wedge
\ldots \wedge dx_{i_k}.
\eeas
On $\d M$, this decomposition does not depend on the choice of coordinates,
and therefore the boundary conditions $\omega_N=$ and $\omega_T=0$ make
sense for $k$-forms, and they do play a role in natural problems.

A version of the Hodge decomposition suitable for our needs has been given by
Iwaniec, Scott, and Stroffolini --- they, as well as Schwarz, proved several
different decompositions with different sets of boundary conditions.
The one we need is from
\cite[Remark 5.1 and Theorem 5.7]{ISS}, improved with arguments from
\cite[Lemma 2.4.11]{Sch} for higher order Sobolev spaces. In what follows,
$\HH^\ell_N(M)$ is the space of harmonic forms on $M$ with normal part
vanishing on $\d M$.

\begin{proposition}[Hodge decomposition with boundary]\la{Pro1}
Let $M$ be a compact $m$-dimensional manifold with smooth boundary.
For $1<p<\infty$, $k\in\N_0$ and every $\ell\in\{0,\ldots,m\}$, the space
$W^{k,p}\Lambda^\ell(M)$ decomposes as a direct sum
\[
  W^{k,p}\Lambda^\ell(M)=dW^{k+1,p}\Lambda^{\ell-1}(M)\oplus d^*W^{k+1,p}_N
  \Lambda^{\ell+1}(M)\oplus\HH^\ell_N(M).
\]
Correspondingly, any $\omega\in W^{k,p}\Lambda^\ell(M)$ can be decomposed as
\[
  \omega=d\a+d^*\b+h\qquad\mbox{ with $\b_N=0$ and $h_N=0$.}
\]
The forms $\a,\b,h$ are uniquely determined  under the further conditions
\[
  \a\in d^*W^{k+2,p}_N\Lambda^\ell(M),\qquad \b\in dW^{k+2,p}\Lambda^\ell(M).
\]
There is a constant $c$ depending on $k$, $p$, and $M$ only such that
\[
  \|\a\|_{W^{k+1,p}(M)}+\|\b\|_{W^{k+1,p}(M)}+\|h\|_{W^{k,p}(M)}\le\|\omega\|_{W^{k,p}(M)}.
\]
If $1\le k\le m$, $m\ge2$, and $M=B^m$, we always have $h\equiv0$.
\end{proposition}

The case $M=B^m$ mentioned in the last line is not explicitly
discussed in the sources mentioned above. We can prove it as follows.
For $1\le k\le m$, by Poincar\'e's Lemma together with the Hodge
isomorphism (see \cite[Theorem 2.6.1]{Sch} for details), we have
$\HH^k_N(B^m)=\{0\}$, which means $h\equiv0$.\qed

Hodge theory on manifolds with boundary has been developed to quite some
extent. Both \cite{ISS} and \cite{Sch} are excellent references.
A closely related mathematical fact is Gaffney's inequality. We state it
in a version that combines \cite[Thms.\ 4.8 and 4.11]{ISS}
and the considerations
we just made for the special case $M=B^m$.

\begin{proposition}[Gaffney's inequality]\la{Gaff}
  Let $M$ be a compact $m$-dimensional manifold with smooth boundary,
  $1<p<\infty$, $0\le k\le m-1$. For every $k$-form $\omega$ on $M$
  with $\omega$, $d\omega$ and $d^*\omega$ in $L^p$ and $\omega_N=0$ or
  $\omega_T=0$ on $\d M$, we have $\omega\in W^{1,p}$ (in the sense that its full
  covariant derivative is also in $L^p$) with the estimate
  \[
  \|\omega\|_{W^{1,p}(M)}\le c(\|\omega\|_{L^p(M)}+\|d\omega\|_{L^p(M)}
  +\|d^*\omega\|_{L^p(M)}).
  \]
  The constant $c$ depends on $M$ and $p$ only.

  Moreover, if $M=B^m$ and $1\le k\le m-1$, the simpler form
  \[
  \|\omega\|_{W^{1,p}(M)}\le c(\|d\omega\|_{L^p(M)}
  +\|d^*\omega\|_{L^p(M)})
  \]
  holds.
\end{proposition}

\section{The setting of $2$-gauge theory}\la{Sec:3}

A {\em crossed module\/} is $(G,H,t,\a)$, where $G$ and $H$ are groups,
and $t:H\to G$ and $\a:G\to\Aut(H)$ are homomorphisms satisfying
{\em $G$-equivariance\/} of $t$,
\eq{cm1}
  t(\a(g)(h))=gt(h)g^{-1}
\eeq
for all $g\in G$, $h\in H$, and the {\em Peiffer identity}, 
\eq{cm2}
  \a(t(h_1))(h_2)=h_1h_2h_1^{-1}
\eeq
for all $h_1,h_2\in H$. If $G,H$ are Lie groups and $t,\a$ are Lie group
homomorphisms, then $(G,H,t,\a)$ is called a {\em Lie crossed module}.

Given a Lie crossed module, we can linearize everything and get Lie
algebras $\gg$ and $\hh$ with Lie algebra homomorphisms $\tu:\hh\to\gg$
and $\au:\gg\to\aut(\hh)$. They satisfy the linearized versions of the
identities above,
\eq{dcm1}
  \tu(\au(x)(\xi))=[x,\tu(\xi)]
\eeq
for all $x\in\gg$, $\xi\in\hh$,
and
\eq{dcm2}
  \au(\tu(\xi))(\nu)=[\xi,\nu]
\eeq
for all $\xi,\nu\in\hh$. Such a structure $(\gg,\hh,\tu,\au)$ is called 
a {\em differential crossed module.}

Besides the $\gg$-action on $\hh$ via $\au$, we also have a $G$-action on
$\hh$, induced by the action via $\a$ on $H$ and also denoted by $\a$.
For later use, we
assume that $\hh$ is equipped with a norm that is $G$-invariant under
the action described by $\a$.
From \re{cm1} and \re{cm2}, we infer the ``mixed relations''
(cf.\ \cite[Section 2.1.1]{MM})
\eq{mr1}
  \tu((\a(g)(\xi))=g\tu(\xi)g^{-1}
\eeq
for all $g\in G$, $\xi\in\hh$, and
\eq{mr2}
  \a(t(h))(\xi)=h\xi h^{-1}
\eeq
for all $h\in H$, $\xi\in\hh$. (We can always pretend working in matrix Lie
algebras, and therefore write $h\xi h^{-1}$ instead of $\ad_h(\xi)$.)

Assume we are given a Lie crossed module $\GG:=(G,H,t,\a)$ and a manifold $M$.
The trivial principal $\GG$-$2$-bundle over $M$ is just the product
$M\times G\times H$ equipped with the homeomorphisms $t$ and $\a$.
A {\em $2$-connection\/} on that $2$-bundle is given by a pair $(A,B)$ of a
$\gg$-valued $1$-form $A$ and an $\hh$-valued $2$-form $B$ on $M$, where
for the moment we assume them to be smooth. For $(A,B)$ to represent a
$2$-connection, we further require the {\em vanishing fake curvature condition}
\[
  dA+A\wedge A-\tu(B)=0.
\]

The notation is to be understood as follows. Let $\Lambda^k(U,\kk)$ be the
vector space of all $\kk$-valued $k$-forms on $U\subseteq\R^m$, where
$\kk$ is any matrix Lie algebra. Every $V\in\Lambda^k(U,\kk)$ can
be written as $\sum_iV^iX_i$, where $X_i\in\kk$ (they need not form a
basis), and the $V^i$ are scalar $k$-forms. Similarly, 
$W\in\Lambda^\ell(U,\kk)$ equals $\sum_jW^jX_j$. Then we write
\eas
  V\wedge W&:=&\sum_{i,j}V^i\wedge W^jX_iX_j,\\
  {} [V\wedge W]&:=&\sum_{i,j}V^i\wedge W^j[X_i,X_j].
\eeas

For any $2$-connection $(A,B)$, we define $F_A$ and the {\em $2$-curvature\/}
$Z_{A,B}$ by
\eas
  F_A&:=&dA+A\wedge A,\\
  Z_{A,B}&:=&dB+\au(A)\wedge B,
\eeas
Basic facts about the curvatures are the two Bianchi identities
\eas
  dF_A+A\wedge F_A&=&0,\\
  d_{A,B}Z+\au(A)\wedge Z_{A,B}&=&0,
\eeas
and their consequence
\[
  \tu(Z_{A,B})=0.
\]

A {\em $2$-gauge transformation\/} is given by a function $g:U\to G$ and a
$\hh$-valued $1$-form $\chi$ on $M$. They transform a $2$-connection
$(A,B)$ to another $2$-connection $(A',B')$ via
\eas
  A'&=&g^{-1}Ag+g^{-1}dg-\tu(\chi),\\
  B'&=&\a(g^{-1})(B)-\au(A')\wedge\chi-d\chi-\chi\wedge\chi.
\eeas
We write $(A',B')=:(g,\chi)^*(A,B)$ and find
\[
  (g',\chi')^*(g,\chi)^*(A,B)=(gg',\a(g')(\chi)+\chi')^*(A,B),
\]
which means that the group of $2$-gauge transformations is that of
functions with values in the semi-direct
product $G\ltimes_\a\hh$ acting from the right on the $2$-connections.

The remarkable thing about the $2$-curvature $Z_{A,B}$ is its covariance under
$2$-gauge transformations $(g,\chi)$. Those transform $Z_{A,B}$ according to
to
\[
  Z_{A',B'}=\a(g^{-1})(Z_{A,B}),
\]
while $F_A$ does not transform nicely. Since the norm on $\hh$ is assumed to be
$G$-invariant via $\a$, this implies that
\[
  YM_2(A,B):=\int_U|Z_{A,B}|^2\,dx
\]
is $2$-gauge invariant (and conformally invariant on $\R^6$). Therefore it is
a promising candidate for a ``higher'' variant of the Yang-Mills functional.

It is well known that $\tu(Z_{A,B})=0$ holds because of Bianchi's identities.
This gives the theory the abelian flavor we already mentioned,
since $\Ker\tu$ is an
abelian subalgebra of $\hh$, which is seen immediately from \re{dcm2}
since $[\xi,\nu]=\au(\tu(\xi))(\nu)=\au(0)(\nu)=0$ for all $\xi\in\Ker\tu$
and all $\nu\in\hh$.

\section{Canonical $2$-gauges in $2$-gauge theory}\la{Sec:4}

A basic aim in gauge theory is to control natural quantities like norms of
connections by gauge invariant quantities, after applying a suitable gauge
transformation. A good example is Uhlenbeck's theorem discussed in the
introduction.
Its proof quite nontrivial, and the smallness condition cannot be entirely
removed. Compared to that, the
corresponding $2$-gauge theorem for $2$-connections is much simpler.

\begin{theorem}[canonical $2$-gauges for $2$-connections]\la{2-ga-th}
  Assume we are given a Lie crossed module $(G,H,t,\a)$ where $G$ is
  a compact Lie group. Assume $3\le m\le6$ and that
  $(A,B)\in W^{2,2}\Lambda^1(B^m,\gg)\times W^{1,2}\Lambda^2(B^m,\hh)$
  represents a $2$-connection of the trivial $2$-bundle associated with
  $(G,H,t,\a)$ over $B^m$. Then there is a $2$-gauge transformation
  $(g,\chi)\in W^{3,2}(B^m,G)\times W^{2,2}\Lambda^1(B^m,\hh)$ such that
  $(A',B'):=(g,\chi)^*(A,B)$ satisfies
  \[
    A'=0,\qquad \tu(B')=0,\qquad d^*B'=0,\qquad (B'_N)_{|\d B^m}=0,
  \]
  and its norm is controlled by the $2$-curvature,
  \[
    \|B'\|_{W^{1,2}(B^m)}\le c\|Z_{0,B'}\|_{L^2(B^m)}=c\|Z_{A,B}\|_{L^2(B^m)}\,.
  \]
  The gauge transformation obeys the estimates
  \eas
    \|dg\|_{W^{2,2}(B^m)}&\le&c\|A\|_{W^{2,2}(B^m)},\\
    \|\chi\|_{W^{2,2}(B^m)}&\le&c(\|A\|_{W^{2,2}(B^m)}+\|A\|_{W^{2,2}(B^m)}^3
      +\|B\|_{W^{1,2}(B^m)}+\|B\|_{W^{1,2}(B^m)}^{3/2}).
  \eeas
  The $2$-connection $(A',B')=(0,B')$ is unique up to a constant gauge
  transformation, i.e.\ up to a $2$-gauge transformation $(g_0,0)$ with
  some $g_0\in G$.
\end{theorem}

{\bf Proof.} Since $G$ is compact, $\gg$ is semisimple.
The image $\tu(\hh)$ of the Lie algebra homomorphism $\hh$ is a Lie subalgebra
of $\gg$. Even better, it is an ideal in $\gg$,
$[\tu(\hh),\gg]\subseteq\tu(\hh)$ because of \re{dcm1}. Now, for any ideal
in a semisimple Lie algebra, the Lie algebra is the direct sum of
the ideal and its orthogonal complement with respect to the Killing form.
In our case, this also means that $\tu(\hh)^\bot$ is a Lie subalgebra of $\gg$.

Fix a right inverse $\tu_{-1}:\tu(\hh)\to\hh$ of $\tu$ for which
$\tu\circ\tu_{-1}$ is the identity of $\tu(\hh)$. Decompose
$A=A^\top+A^\bot$ according to the direct sum
$\tu(\hh)\oplus\tu(\hh)^\bot$. Under the
$2$-gauge transformation
$(e,\chi_1):=(e,\tu_{-1}(A^\top))$, $(A,B)$ transforms to
\[
  (A_1,B_1)=(A^\bot,B-\au(A^\bot)\wedge\tu_{-1}(A^\top)
  -\tu_{-1}(dA^\top+A^\top\wedge A^\top)).
\]
Since $A^\bot$ takes its values in $\tu(\hh)^\bot$, so do $dA^\bot$ and
$A^\bot\wedge A^\bot$, the latter because $\tu(\hh)^\bot$ is a Lie subalgebra.
Hence $F_{A_1}$ is a $\tu(\hh)^\bot$-valued $2$-form. But it is also
$\tu(\hh)$-valued because of $\tu(B_1)=F_{A_1}$. This means that $F_{A_1}=0$.

Now that we know $F_{A_1}=0$, we have $\|F_{A_1}\|_{L^3(B^m)}=0$, and
of course $A_1\in W^{1,3}\cap L^6$ because of Sobolev's embeddings
$W^{2,2}\hookrightarrow W^{1,3}\hookrightarrow L^6$. Hence
Uhlenbeck's theorem gives us a gauge transformation $g_2\in W^{3,2}(B^m,G)$
such that $g_2^*A_1=0$ on $B^m$ because of
$\|g_2^*A_1\|_{W^{1,3}(B^m)}\le c\|F_{A_1}\|_{L^3(B^m)}=0$. This means we can apply
the $2$-gauge transformation $(g_2,0)$ to $(A_1,B_2)$ to find that $(A,B)$
is $2$-gauge equivalent to
\[
  (A_2,B_2)=(0,\a(g_2^{-1})(B_1)).
\]

Apart from having $A_2=0$, we also know that $B_2$ is quite simple (like $B_1$,
in fact) because we have $\tu(B_2)=F_{A_2}=0$, hence $B_2$ takes its values
in $\Ker\tu$, and we have seen above that on $\Ker\tu\subseteq\hh$,
the Lie bracket of $\hh$ vanishes.
Remember that $Z$ always takes its values in
$\Ker\tu$. And now that we have $A_2=0$, we have $Z_{A_2,B_2}=dB_2$.

The question of finding a good $2$-gauge at this stage is reduced to a
completely linear problem. We have $Z_{A_2,B_2}=dB_2$ and
\[
  (e,\chi_3)^*(0,B_2)=(0,B_2-d\chi_3)
\]
if we assume that also $\chi_3$ takes its values in $\Ker\tu$. But now that
everything is linear and without Lie brackets, the gauge problem reduces
simply to a question in Hodge theory. We use the Hodge decomposition from
Proposition~\ref{Pro1}. Having $B_2\in W^{2,2}\Lambda^2(B^m,\Ker\tu)$,
we find unique forms $a\in d^*W^{4,2}\Lambda^1(B^m,\Ker\tu)$ and
$b\in dW^{4,2}\Lambda^1(B^m,\Ker\tu)$ satisfying $b_N=0$ on $\d B^m$
such that $B_2=da+d^*b$. 
We then choose $\chi_3:=a$ and find
\[
  (A',B'):=(e,\chi_3)^*(0,B_2)=(0,d^*b),
\]
which proves the existence of a suitable gauge.

Concerning the estimate of the norm of $B'$ by that of $Z$,
we note that $b_N=0$ implies $d^*b_N=0$
on $\d B^m$. We therefore have $d^*B'=0$, $dB'=Z_{0,B'}$, and $(B'_N)_{|\d B^m}=0$,
which means we can use Gaffney's inequality Proposition \ref{Gaff} to estimate
\[
  \|B'\|_{W^{1,2}(B^m)}\le c(\|dB'\|_{W^{1,2}(B^m)}+\|d^*B'\|_{W^{1,2}(B^m)})
  \le c\|Z_{0,B'}\|_{L^2(B^m)}.
\]

What remains to be shown are estimates for $g$ and $\chi$, which are given by
composition of $(e,\chi_1)$, $(g_2,0)$, and $(e,\chi_3)$,
\eas
  g&=&g_2,\\
  \chi&=&\a(g_2)(\tu_{-1}(A^\top))+\g.
\eeas
By an easy consequence of Uhlenbeck's theorem (that is, bootstrapping
and using the equation $dg_2=-A_1g_2$ as in \cite[Lemma A.8]{We}),
we have
\[
  \|dg_2\|_{W^{2,2}(B^m)}\le c\|A_1\|_{W^{2,2}(B^m)}\le c\|A\|_{W^{2,2}(B^m)}.
\]
And using the trivial estimates $\|A^\top\|_{W^{2,2}(B^m)}\le \|A\|_{W^{2,2}(B^m)}$
and $\|g_2\|_{W^{3,2}(B^m)}\le c+\|dg_2\|_{W^{2,2}(B^m)}$,
we estimate
\eas
  \|\a(g_2)(\tu_{-1}(A^\top))\|_{W^{2,2}}
  &\le&c(\|g_2\|_{W^{2,3}}\|A\|_{L^6}+\|g_2\|_{W^{1,6}}\|A\|_{W^{1,3}}+\|g_2\|_{L^\infty}
    \|A\|_{W^{2,2}})\\
  &\le&c(\|g\|_{W^{3,2}}\|A\|_{W^{2,2}})\\
  &\le&c(\|A\|_{W^{2,2}}+\|A\|_{W^{2,2}}^2),
\eeas
with all norms to be taken on $B^m$. Combining that with the estimate from
the Hodge decomposition,
\eas
  \|a\|_{W^{2,2}}&\le&c\|B_2\|_{W^{1,2}}\\
  &=&c\|\a(g_2^{-1})(B-\au(A)\wedge\tu_{-1}(A^\top)
    -\tu_{-1}(dA^\top+A^\top\wedge A^\top))\|_{W^{1,2}}\\
  &\le&c(\|B\|_{W^{1,2}}+\|A\|_{W^{1,3}}\|A\|_{L^6}+\|dA\|_{W^{1,2}})\\
  &&{}+c\|dg_2\|_{L^6}(\|B\|_{L^3}+\|A\|_{L_6}^2+\|dA\|_{L^3})\\
  &\le&c(\|B\|_{W^{1,2}}+\|B\|_{W^{1,2}}^2+\|A\|_{W^{2,2}}+\|A\|_{W^{2,2}}^3),
\eeas
we have proven the estimates for $g$ and $\chi$.

Now we turn to the uniqueness of $B'$. Assume that we have a
``canonical gauge'' $(0,B'')$ of $(A,B)$ with the same properties as
$(0,B')$. Then there exists a $2$-gauge transformation, again denoted by
$(g,\chi)$, such that $(0,B'')=(g,\chi)^*(0,B')$, which means
\ea{u1}
0&=&g^{-1}dg-\tu(\chi),\\
\la{u2}
B''&=&\a(g^{-1})(B')-d\chi-\chi\wedge\chi.
\eea
Observe that \re{u1} implies $g^{-1}dg\in\tu(\hh)$ almost everywhere, which
means that $g$ is of the form $g_0t(h)$ for some constant $g_0\in G$ and some
$h\in W^{3,2}(B^m,H)$. And $\a(g_0^{-1})(B')\in\Ker\tu$ because $\tu(B')=0$
implies $\tu(\a(g_0^{-1})(B'))=g_0^{-1}\tu(B')g_0=0$ via \re{mr1}. Since the
adjoint action of $H$ on the abelian subalgebra $\Ker\tu$ of $\hh$ is trivial,
using \re{mr2}, we find
\[
  \a(g^{-1})(B')=\a(t(h^{-1}))(\a(g_0^{-1})(B'))=h^{-1}\a(g_0^{-1})(B')h
  =\a(g_0^{-1})(B'),
\]
Now this means that \re{u2} simplifies to
\eq{u4}
B''=\a(g_0^{-1})(B')-d\chi-\chi\wedge\chi=:\a(g_0^{-1})(B')-\nu.
\eeq
Applying $d$ to \re{u4}, we have
\[
dB''=\a(g_0^{-1})(dB')-d\nu,
\]
which we compare to an equation using the transformation behavior of $Z$
and $A'=A''=0$,
\[
dB''=Z_{0,B''}=\a(g^{-1})(Z_{0,B'})=\a(g_0^{-1})(Z_{0,B'})=\a(g_0^{-1})(dB'),
\]
where the third ``$=$'' is justified as above using $\tu(Z_{0,B'})=0$.
Comparing the last two equations, we find $d\nu=0$.

We can also apply $d^*$ to \re{u4} to find $d^*\nu=0$ because of $d^*B''=0$
and $d^*B'=0$. And similarly, we observe $\nu_N=0$ on $\d B^m$ since
$B'_N=0$ and $B''_N=0$ on $\d B^m$. Now we know $d\nu=0$ and $d^*\nu=0$
on $B^m$, and $\nu_N=0$ on $\d B^m$, which together imply $\nu=0$, again by
Gaffney's inequality. Then
\re{u4} reads $B''=\a(g_0^{-1})(B')$, which is the asserted uniqueness
of $B'$ modulo constant gauge transformations.\qed

{\bf Remarks.}

{\bf (1)} We have formulated our gauge theorem under the minimal regularity
assumptions on $A$ and $B$. If both $A$ and $B$ have more regularity, we
will have more regularity of $B'$, $dg$, and $\chi$, by the same
proof combined with some iterated estimates. This way, we can easily
formulate $W^{k,p}$- and $C^{k,\a}$-versions of the gauge theorem. For example,
if $A\in W^{k+1,2}$, $B\in W^{k,2}$ for some $k\ge1$, we can choose the
gauge transformation $(g,\chi)$ in $W^{k+2,2}\times W^{k+1,2}$ and control
$B'$ in $W^{k,2}$.

{\bf (2)} If $(0,B')$ is in the ``canonical'' gauge and $2$-Yang-Mills, then
it is stationary for the $2$-YM functional among all connections in canonical
gauge. And since $Z_{0,B'}=dB'$ for those connections, the Euler-Lagrange
equation for that problem is $d^*dB'=0$. Together with $d^*B'=0$ (and
the boundary condition for $B'$), we find that $\Delta B'=0$ is equivalent to
the $2$-Yang-Mills equation for all connections in canonical gauge. This means
that transforming to the canonical gauge, the (nonlinear)
$2$-Yang-Mills equation reduces to the (linear) Laplace equation.

In contrast to this, the Euler-Lagrange equation for the $2$-Yang-Mills energy
is more difficult to write down in general gauge, because the Euler-Lagrange
equation for $\int|dB+\au(A)\wedge B|^2\,dx$ has to be derived under
the side condition $F_A-\tu(B)=0$. The full system looks like
\eas
  d_A^*d_AB&=&\tu^*(\l),\\
  d^*\l+[A\ecke\l]&=&-\au^*(B\ecke d_AB),\\
  F_A&=&\tu(B),
\eeas
where $\l\in\Lambda^2(B^m,\gg)$ is an unknown Lagrange multiplier, and
``$\ecke$'' denotes suitable contractions of forms.

{\bf (3)} The previous remark has an interesting consequence, which generalizes
the classical fact that (anti-)selfdual connections in $4$ dimensions are
Yang-Mils. For any $2$-connection $(A,B)$ over $B^6$,
the $2$-curvature $Z_{A,B}$ is (anti-)selfdual if $*Z_{A,B}=\pm Z_{A,B}$. We then
call also $(A,B)$ an (anti-)selfdual $2$-connection. For a $2$-connection
$(0,B')$ in canonical gauge, (anti-)selfduality means $dB=\pm*dB$.
We then have $d^*dB'=\pm*d*(dB')=\pm*d**dB'=0$. Again, together with $d^*B'=0$,
we find $\Delta B'=0$. Hence any
(anti-)selfdual $2$-connection in canonical gauge is also $2$-Yang-Mills.
And since both the (anti-)selfduality and the $2$-Yang-Mills functional (and
hence its equations) are invariant under $2$-gauge transformations, this
proves:

\begin{corollary}\la{2-Kor}
  Every (anti-)selfdual $2$-connection $(A,B)$ over $B^6$ is also
  $2$-Yang-Mills.
\end{corollary}

\section{The setting of $3$-gauge theory}\la{Sec:5}

There is a notion of $3$-gauge theory which is based on Lie $2$-crossed modules.
It has been developed systematically by S\"amann and Wolf \cite{SW}.
We refer to \cite{Wa} for a concise presentation of the algebraic aspects of
the local theory.
Which is described using a complex of Lie groups
\[
  L\stackrel{\tau}{\longrightarrow}H\stackrel{t}{\longrightarrow}G
\]
(with $t\circ\tau\equiv e$). We also need homomorphisms $\a:G\to\Aut(H)$
and $\b:G\to\Aut(L)$, with respect to which $t$ and $\tau$ are again
$G$-equivariant. That means \re{cm1}
holds, and also $\tau(\b(g)(\ell))=\a(g)(\tau(\ell))$ for all
$g\in G,\ell\in L$. The Peiffer identity
is now replaced by a {\em Peiffer lifting\/} which is a smooth function
$\{\,\cdot\,,\,\cdot\,\}:H\times H\to L$ that is $G$-equivariant in the sense
that
\[
  \b(g)(\{h,k\})=\{\a(g)(h),\a(g)(k)\}
\]
for all $g\in G$, $h,k\in H$. Moreover, it must satisfy the relations
\ea{P5}
\tau(\{h,k\})&=&hkh^{-1}\a(t(h))(k^{-1})=:\<h,k\>,\nn\\
lml^{-1}m^{-1}&=&\{\tau(l),\tau(m)\},\nn\\
\{hj,k\}&=&\{h,jkj^{-1}\}\a(t(h))(\{j,k\}),\nn\\
\{h,jk\}&=&\{h,j\}\{h,k\}\{\<h,k\>^{-1},\a(t(h))(j)\},\nn\\
\{\tau(l),h\}\{h,\tau(l)\}&=&l\b(t(h))(l^{-1})
\eea
for all $h,j,k\in H$ and $l,m\in L$.

Correspondingly, a {\em differential $2$-crossed module\/} is described by
a complex of Lie algebras
\[
  \ll\stackrel{\su}{\longrightarrow}\hh\stackrel{\tu}{\longrightarrow}\gg
\]
with $t\circ\tau\equiv 0$ and a Peiffer lifting
$\{\,\cdot\,,\,\cdot\,\}:H\times H\to L$ which
is $\gg$-equivariant in the sense that
\[
  \bu(a)(\{u,v\})=\{\au(a)(u),\au(a)(v)\}
\]
for all $a\in\gg$, $u,v\in\hh$, where here $\au:\gg\to\aut(\hh)$ and
$\bu:\gg\to\aut(\ll)$ are $\gg$-invariant Lie algebra homomorphisms.
The relations for the Peiffer lifting in their linearized versions read
\ea{p1}
\su(\{u,v\})&=&[u,v]-\au(\tu(u))(v),\\
\la{p2}
[x,y]&=&\{\su(x),\su(y)\},\\
\{[u,v],w\}&=&\au(\tu(u))(\{v,w\})+\{u,[v,w]\}-\au(\tu(v))(\{u,w\})
  -\{v,[u,w]\},\nn\\
\{u,[v,w]\}&=&\{\su(\{u,v\}),w\}-\{\su(\{u,w\}),v\},\nn\\
\la{p5}  
\{\su(x),u\}+\{u,\su(x)\}&=&-\bu(\tu(u))(x)
\eea
for $u,v,w\in\hh$ and $x,y\in\ll$.

Between \re{P5} and \re{p5}, there is another ``mixed relation'' concerning
the $G$-operation on $\ll$. We rewrite \re{P5} as
\[
  \b(t(h))(\ell)=\ell\{\tau(\ell^{-1}),h\}\{h,\tau(\ell^{-1})\}.
\]
Using $\{e_H,h\}=\{h,e_H\}=e_L$ and letting $\ell=\exp(rx)$, we can
differentiate at $r=0$ to find
\ea{mr3}
\b(t(h))(x)&=&x\{\tau(e_L),h\}\{h,\tau(e_H)\}-e_L\{\su(x),h\}\{h,\tau(e_L)\}
-e_L\{\tau(e_L),h\}\{h,\su(x)\}\nn\\
&=&x-\{\su(x),h\}-\{h,\su(x)\}
\eea
for all $h\in H,x\in\ll$, in a calculation that involves three different
$\{\,\cdot\,,\,\cdot\,\}$
living on $H\times H$, $H\times\hh$, and $\hh\times H$. The mixed relation
from the $G$-equivariance of $\tau$ is
\[
\su(\b(g)(x))=\a(g)(\su(x))
\]
for all $g\in G, x\in\ll$.

It was proven in \cite[Proposition 2.2]{Wa} that for $1$-forms
$\chi,\eta\in\Lambda^1(U,\hh)$, \re{p1} implies
\[
\au(\tu(\chi))(\eta)=[\chi\wedge\eta]-\su(\{\chi\wedge\eta\}),
\]
and in particular
\eq{p1a}
\chi\wedge\chi=\frac12\,[\chi\wedge\chi]=\frac12\,\au(\tu(\chi))(\chi)
+\frac12\,\su(\{\chi\wedge\chi\}).
\eeq

A $3$-connection on the trivial $3$-bundle over $U$ is a triple
\[
(A,B,C)\in\Lambda^1(U,\gg)\times\Lambda^2(U,\hh)\times\Lambda^3(U,\ll)
\]
satisfying two ``fake curvature conditions''
\ea{fc1}
dA+A\wedge A&=&\tu(B),\\
\la{fc2}
dB+\au(A)\wedge B&=&\su(C).
\eea
A {\em $3$-gauge transformation\/} is a triple $(g,\chi,\l)$ of a function
$g:U\to G$, an $\hh$-valued $1$-form $\chi\in\Lambda^1(U,\hh)$, and
an $\ll$-valued $2$-form $\l\in\Lambda^2(U,\ll)$, acting on $3$-connections
via
\eas
A'&=&g^{-1}Ag+g^{-1}dg-\tu(\chi),\\
B'&=&\a(g^{-1})(B)-A'\wedge\chi-d\chi-\chi\wedge\chi-\su(\l),\\
C'&=&\b(g^{-1})(C)-d\l-\bu(A')\wedge\l+\{B'\wedge\chi\}
  +\{\chi\wedge\a(g^{-1})(B)\}+\{\su(\l)\wedge\chi\}.
\eeas
In particular, any $(g,0,0)$ acts via
\eas
A'&=&g^{-1}Ag+g^{-1}dg,\\
B'&=&\a(g^{-1})(B),\\
C'&=&\b(g^{-1})(C),
\eeas
while $(e,\chi,0)$ acts via
\eas
A'&=&A-\tu(\chi),\\
B'&=&B-\au(A')\wedge\chi-d\chi-\chi\wedge\chi,\\
C'&=&C+\{B'\wedge\chi\}+\{\chi\wedge B'\},
\eeas
and $(e,0,\l)$ via
\eas
A'&=&A,\\
B'&=&B-\su(\l),\\
C'&=&C-d\l-\bu(A')\wedge\l.
\eeas
The $3$-curvature, transforming naturally under any of these, is an
$\ll$-valued $4$-form given by
\[
Y_{A,B,C}:=dC+\bu(A)\wedge C+\{B\wedge B\}.
\]
In particular, the $3$-Yang-Mills functional
\[ YM_3(A,B,C):=\int_U|Y_{A,B,C}|^2\,dx \]
is invariant under all $3$-gauge transformations, and conformally invariant
if $m=8$. Of course, this uses $Y_{A',B',C'}=\b(g^{-1})(Y_{A,B,C})$ for
$(A',B',C'):=(g,\chi,\l)^*(A,B,C)$, and the asserted invariance of $YM_3$
can only hold if we have assumed $G$-invariance (via $\b$) of the norm we
have chosen on $\ll$.

\section{Canonical $3$-gauges in $3$-gauge theory}\la{Sec:6}

Here is our analogue of Theorem \ref{2-ga-th} for $3$-connections.

\begin{theorem}[canonical $3$-gauges for $3$-connections]\la{3-ga-th}
  Assume we are given a Lie $2$-crossed module $(G,H,L,t,\tau,\a,\b)$ where $G$
  is a compact Lie group. Assume $4\le m\le8$ and that
  $(A,B,C)\in W^{3,2}\Lambda^1(B^m,\gg)\times W^{2,2}\Lambda^2(B^m,\hh)\times
  W^{1,2}\Lambda^3(B^m,\ll)$
  represents a $3$-connection of the trivial $3$-bundle associated with
  $(G,H,L,t,\tau,\a,\b)$ over $B^m$. Then there is a $3$-gauge transformation
  $(g,\chi,\l)\in W^{4,2}(B^m,G)\times W^{3,2}\Lambda^1(B^m,\hh)\times
  W^{2,2}\Lambda^2(B^m,\ll)$ such that
  $(A',B',C'):=(g,\chi,\l)^*(A,B)$ satisfies
  \[
    A'=0,\quad B'=0,\quad \su(C')=0,\quad d^*C'=0,\quad (C'_N)_{|\d B^m}=0,
  \]
  and its norm is controlled by the $3$-curvature,
  \[
    \|C'\|_{W^{1,2}(B^m)}\le c\|Y_{0,0,C'}\|_{L^2(B^m)}=c\|Y_{A,B,C}\|_{L^2(B^m)}\,.
  \]
  The gauge transformation obeys the estimates 
  \eas
    \|dg\|_{W^{3,2}(B^m)}&\le&c\|A\|_{W^{3,2}(B^m)},\\
    \|\chi\|_{W^{3,2}(B^m)}&\le&c(\|A\|_{W^{3,2}(B^m)}+\|A\|_{W^{3,2}(B^m)}^4
    +\|B\|_{W^{2,2}(B^m)}+\|B\|_{W^{2,2}(B^m)}^2),\\
    \|\l\|_{W^{2,2}(B^m)}&\le&c(\|A\|_{W^{3,2}(B^m)}+\|A\|_{W^{3,2}(B^m)}^4
    +\|B\|_{W^{2,2}(B^m)}+\|B\|_{W^{2,2}(B^m)}^2\\
    &&\qquad+\|C\|_{W^{1,2}(B^m)}+\|C\|_{W^{1,2}(B^m)}^{4/3}).
  \eeas
\end{theorem}

{\bf Proof.}
Exactly as in the proof of Theorem \ref{2-ga-th}, we make gauge transformations
that transform $(A,B,C)$ to $(0,B_2,C_2)$ with $B_2$ a $\Ker\tu$-valued
$1$-form. Now decompose $B_2=B_2^\top+B_2^\bot$ according to the direct sum
$\su(\ll)\oplus\su(\ll)^\bot$.
Assume again we have chosen some fixed right inverse $\su_{-1}:\su(\ll)\to\ll$
of $\su$. Apply the gauge transformation
\[
(A_3,B_3,C_3):=(e,0,\su_{-1}(B_2^\top))^*(0,B_2,C_2)
\]
and find
\eas
A_3&=&0,\\
B_3&=&B_2^\bot,\\
C_3&=&C_3-\su_{-1}(dB_2^\top).
\eeas
Now that we know $B_3$ is an $\su(\ll)^\bot$-valued $1$-form, we can
Hodge-decompose $B_3$ in the space of such forms, which means we find
unique forms $a\in d^*W^{5,2}\Lambda^1(B^m,\su(\ll)^\bot)$ and
$b\in dW^{5,2}\Lambda^2(B^m,\su(\ll)^\bot)$ satisfying $b_N=0$ on $\d B^m$
such that $B_2=da+d^*b$.
Having that, we perform the gauge transformation
\[
(A_4,B_4,C_4):=(e,a,0)^*(0,B_3,C_3)
\]
with the result
\eas
A_4&=&0,\\
B_4&=&d^*b-a\wedge a,\\
C_4&=&C_2+\{B_2\wedge a\}+\{a\wedge B_2\}.
\eeas
Depending on the structure of $\hh$, it may well be that $a\wedge a=0$ holds
automatically. But we can transform it away, anyway. From
$\tu(a)=0$ and \re{p1a}, we find
\[
a\wedge a=\frac12\,\su(\{a\wedge a\}),
\]
which means our next step should be the gauge transformation
\[
(A_5,B_5,C_5):=(e,0,-\tf12\{a\wedge a\})^*(0,B_4,C_4),
\]
where here
\eas
A_5&=&0,\\
B_5&=&d^*b,\\
C_5&=&C_4+\tf12d\{a,a\}.
\eeas
Now we have that $B_5$ is a $\su(\ll)^\bot$-valued $2$-form,
which implies that also
$dB$ is $\su(\ll)^\bot$-valued. On the other hand, \re{fc2} (together with
$A_5=0$) implies that $dB_5=Z_{0,B_5}=\su(C_5)$ takes its values in
$\su(\ll)$, which means $dB_5=0$. Since also $d^*B_5=d^*d^*b=0$, and $(B_5)_N=0$,
we now know that actually $B_5=0$ by Gaffney's inequality.

We have reached a situation that parallels that for $(0,B_2)$ in the previous
section. We know that $(A_5,B_5,C_5)=(0,0,C_5)$, and because of $\su(C_5)=
Z_{0,0}=0$, we know that $C_5$ takes its values in $\Ker\su\subset\ll$.
Because of \re{p2}, the Lie subalgebra $\Ker\su$ of $\ll$ is abelian.
We can again perform Hodge decomposition, this time
$C_5=dp+d^*q$, $p\in d^*W^{4,2}\Lambda^2(B^m,\Ker\su)$ and
$q\in dW^{5,2}\Lambda^2(B^m,\Ker\su)$ with $q_N=0$ on $\d B^m$. We let
\[
  (A',B',C'):=(e,0,p)^*(0,0,C_5)=(0,0,d^*q),
\]
which is the transformed $3$-connection as stated in the theorem.
The estimates for the gauge transform put together from those in the
proof are a routine calculation along the lines of the proof of
Theorem~\ref{2-ga-th}.

The uniqueness proof for $C'$ is also similar, but needs a few modifications.
Assume again that we have two canonical gauges $(0,0,C')$ and $(0,0,C'')$
of $(A,B,C)$. Then there exists a $3$-gauge transformation, again denoted by
$(g,\chi,\l)$, such that $(0,0,C'')=(g,\chi,\l)^*(0,0,C')$, meaning
\ea{U1}
\tu(\chi)&=&g^{-1}dg,\\
\la{U2}
\su(\l)&=&-d\chi-\chi\wedge\chi,\\
\la{U3}
C''&=&\b(g^{-1})(C')-d\l+\{\su(\l)\wedge\chi\}. 
\eea
As before, \re{U1} implies $g=g_0t(h)$ with $g_0\in G$ constant.
From \re{mr3}, we have
\[
\b(t(h^{-1}))(X)=X-\{\su(X),h^{-1}\}-\{h^{-1},\su(X)\}=X
\]
for any $k$-form $X$ with values in $\Ker\su$. We can apply this to
$X=\b(g_0^{-1})(C')$ to simplify \re{U3}, finding
\eq{U4}
C''=\b(g_0^{-1})(C')-d\l+\{\su(\l)\wedge\chi\}=:\b(g_0^{-1})(C')-\xi, 
\eeq
and to $X=dC'$ in
\[
dC''=Y_{0,0,C''}=\b(g^{-1})(Y_{0,0,C'})=\b(g_0^{-1})(Y_{0,0,C'})=\b(g_0^{-1})(dC').
\]
Comparing the latter to $d$\re{U4}, we have $d\xi=0$, and from $d^*$\re{U4} and
\re{U4}$\strut_N$, we again have $d^*\xi=0$ on $B^m$ and $\xi_N=0$ on
$\d B^m$. Hence $\xi=0$, which completes the uniqueness proof.\qed

All remarks made about our $2$-gauge theorem apply similarly here. In
particular, we have, by the same reasoning as for Corollary \ref{2-Kor},
a selfduality theorem. In the case of an $8$-dimensional base of the
$3$-bundle, we call a $3$-connection $(A,B,C)$ (anti-)selfdual if
$*Y_{A,B,C}=\pm Y_{A,B,C}$.

\begin{corollary}\la{3-Kor}
  Every (anti-)selfdual $3$-connection $(A,B,C)$ over $B^8$ is also
  $3$-Yang-Mills.
\end{corollary}

\vfill

\begin{center}
\scriptsize
Fakult\"at f\"ur Mathematik, Universit\"at Duisburg-Essen, D-45117 Essen, 
Germany.\\
{\tt andreas.gastel@uni-due.de}
\end{center}

\end{document}